\documentclass[letterpaper,superscriptaddress,nofootinbib,longbibliography,11pt]{revtex4-1}
\usepackage[T1]{fontenc} 
\usepackage{amsmath}    
\usepackage{graphicx}   
\usepackage{verbatim}   
\usepackage{color}      
\usepackage[caption=false]{subfig}
\usepackage{adjustbox}
\usepackage{multirow}
\usepackage[%
hidelinks,
  breaklinks=true,%
  colorlinks=true,%
  linkcolor=blue,anchorcolor=blue,%
  citecolor=blue,filecolor=blue,%
  menucolor=blue,%
  urlcolor=blue]{hyperref}  
\usepackage[separate-uncertainty,retain-explicit-plus,per-mode=symbol,binary-units]{siunitx}
\usepackage{rotate}
\usepackage{rotating}
\usepackage{bm}
\usepackage{paralist}
\usepackage{wrapfig}
\usepackage{caption3}
\usepackage{longtable}
\usepackage{appendix}
\usepackage{vruler}
\usepackage{enumitem}
\usepackage[version=4]{mhchem}
\usepackage{chemarrow}

\usepackage{titlesec}
\titleformat{\section}{\normalfont\fontsize{11}{15}\bfseries\centering}{\thesection.}{1em}{\uppercase}
\titleformat{\subsection}{\normalfont\fontsize{11}{15}\bfseries\centering}{\thesection\,\thesubsection.}{1em}{}
\titlespacing{\section}{0pt}{0.8\baselineskip}{0.5\baselineskip}
\titlespacing{\subsection}{0pt}{0.8\baselineskip}{0.5\baselineskip}
\titlespacing{\subsubsection}{0pt}{0.4\baselineskip}{0.2\baselineskip}

\renewcommand{\thesection}{\arabic{section}}
\renewcommand{\thesubsection}{\Alph{subsection}}

\newcommand{\um}{\mbox{$\mu$m}}

\newcommand{\bb}{\mbox{$\beta\beta$}}

\newcommand{\mbb}{\mbox{$m_{\beta\beta}$}}

\newcommand{\gr}{\mbox{$\gamma$-ray}}
\newcommand{\nue}{\mbox{$\nu_e$}}
\newcommand{\ise}{\mbox{$^{82}$Se}}

\newcommand{\ura}{\ce{^{238}U}}
\newcommand{\tho}{\ce{^{232}Th}}

\DeclareSIUnit\c{\mbox{$c$}}
\DeclareSIUnit\year{yr}
\DeclareSIUnit\electron{\mbox{$e^-$}}
\DeclareSIUnit\inch{in}
\newcommand{\adsurl}[1]{\href{#1}{ADS}} 
\providecommand{\url}[1]{\href{#1}{#1}} 

\usepackage{xspace}
\newcommand{\TMIIm}{\mbox{\emph{Topmetal-II\raise0.5ex\hbox{-}}}\xspace}

\newcommand\snowmass{\begin{center}\rule[-0.2in]{\hsize}{0.01in}\\\rule{\hsize}{0.01in}\\
\vskip 0.1in Submitted to the  Proceedings of the US Community Study\\ 
on the Future of Particle Physics (Snowmass 2021)\\ 
\rule{\hsize}{0.01in}\\\rule[+0.2in]{\hsize}{0.01in} \end{center}}

\begin{document}

\snowmass

\title{The Selena Neutrino Experiment}

\author{A.E.~Chavarria}
\affiliation{Center for Experimental Nuclear Physics and Astrophysics, University of Washington, Seattle, WA 98195, United States}

\author{C.~Galbiati}
\affiliation{Physics Department, Princeton University, Princeton, NJ 08544, United States}
\affiliation{INFN Laboratori Nazionali del Gran Sasso, 67010 Assergi (AQ), Italy}
\affiliation{Gran Sasso Science Institute, 67100 L'Aquila, Italy}

\author{B.~Hernandez-Molinero}
\affiliation{Laboratorio Subterr\'aneo de Canfranc, 22880 Canfranc-Estaci\'on, Spain}

\author{Al.~Ianni}
\affiliation{INFN Laboratori Nazionali del Gran Sasso, 67010 Assergi (AQ), Italy}

\author{X.~Li}
\affiliation{Physics Division, Lawrence Berkeley National Laboratory, Berkeley, CA 94720, United States}

\author{Y.~Mei}
\affiliation{Nuclear Science Division, Lawrence Berkeley National Laboratory, Berkeley, CA 94720, United States}

\author{D.~Montanino}
\affiliation{Dipartimento di Matematica e Fisica ``Ennio De Giorgi'', Universit\`a del Salento, 73100 Lecce, Italy}
\affiliation{INFN Sezione di Lecce, 73100 Lecce, Italy}

\author{X.~Ni}
\affiliation{Center for Experimental Nuclear Physics and Astrophysics, University of Washington, Seattle, WA 98195, United States}

\author{C.~Pe\~na~Garay}
\affiliation{Laboratorio Subterr\'aneo de Canfranc, 22880 Canfranc-Estaci\'on, Spain}
\affiliation{Institute for Integrative Systems Biology {\rm (I$^2$SysBio)}, Valencia, Spain}

\author{A.~Piers}
\affiliation{Center for Experimental Nuclear Physics and Astrophysics, University of Washington, Seattle, WA 98195, United States}

\author{H.~Wang}
\affiliation{Physics and Astronomy Department, University of California, Los Angeles, CA 90095, United States}

\begin{abstract}
Imaging devices made from an ionization target layer of amorphous selenium (aSe) coupled to a silicon complementary metal-oxide-semiconductor (CMOS) active pixel array for charge readout are a promising technology for neutrino physics. The high spatial resolution in a solid-state target provides unparalleled rejection of backgrounds from natural radioactivity in the search for neutrinoless \bb\ decay and for electron neutrino (\nue ) spectroscopy with \ise . In this white paper, we summarize the broad scientific program of a large imaging detector with a 10-ton target of \ise . We review the detector technology, and outline the ongoing research program to realize this experiment.
\end{abstract}

\maketitle

\section{Executive Summary}
\label{sec:summary}

\begin{figure}[t!]
  \begin{center}
    \includegraphics[width=\textwidth]{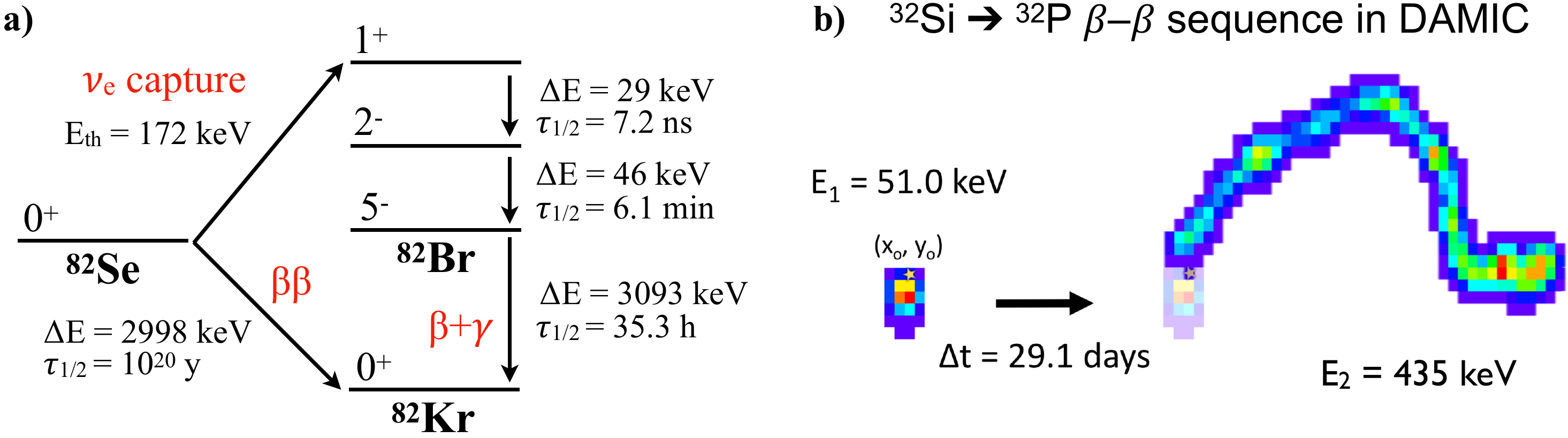}
    \caption{{\bf a)} Diagram showing two natural channels for $^{82}$Se to transmute into $^{82}$Kr. Both $\beta\beta$ decay and $\nu_e$ capture will be detected with high efficiency by Selena for nuclear physics studies. {\bf b)} The identification of a single $^{32}$Si atom by DAMIC at SNOLAB: a sequence of two $\beta$ decays separated in time by many days and with the start points of the electron tracks detected at the same location. Pixels toward the red end of the spectrum collect more charge.}
    \label{fig:selena_principle}
  \end{center}
\end{figure}

The Selena experiment was proposed in Ref.~\cite{Chavarria:2016hxk} to search for the neutrinoless $\beta\beta$ decay of $^{82}$Se with unprecedented sensitivity.  Recently, we expanded its scientific case to include solar neutrino spectroscopy (Sec.~\ref{sec:selena_solar}) and sterile neutrino searches (Sec.~\ref{sec:selena_oscillations}) based on $\nu_e$ capture on $^{82}$Se.  Fig.~\ref{fig:selena_principle}a summarizes the nuclear processes that will be studied by Selena.  Thus, Selena will be a neutrino observatory with a scientific program that spans fundamental neutrino physics and solar astrophysics.

The Selena detector will consist of towers of imaging modules made from $\sim$5\,mm-thick amorphous selenium (aSe)\textemdash isotopically enriched in $^{82}$Se\textemdash deposited on a CMOS active pixel charge sensor (APS).  Free charge liberated by ionizing particles in the aSe is drifted by an applied electric field and collected by the electrodes of CMOS pixels to form high resolution images.  From these images, Selena can $\emph{i)}$~measure the deposited energy, $\emph{ii)}$~ identify the type and number of ionizing particles (electrons, $\alpha$'s, etc.) from the track topologies,  and $\emph{iii)}$~identify radioactive decay sequences by spatio-temporal correlations.  This strategy has been realized to some extent\textemdash in the context of dark matter searches\textemdash by the DAMIC experiment to constrain the activities in the silicon target of $^{32}$Si, and every isotope in the $^{238}$U and $^{232}$Th decay chains~\cite{DAMIC:2020wkw}.  For example, Fig.~\ref{fig:selena_principle}b shows the identification of a $^{32}$Si decay ($Q$-value $=225$\,keV, $\tau_{1/2}=150$\,y) from its spatial correlation with the $\beta$ track of its daughter $^{32}$P ($Q$-value $=1.71$\,MeV, $\tau_{1/2}=14$\,d) many days later.  Note that the start and end points of the $\beta$ track from $^{32}$P decay can be distinguished by the presence of a high density of charge, \emph{i.e.}, the Bragg peak, at the end of the track.  In the context of Selena, this experimental strategy will allow for background-free spectroscopy of $\beta\beta$ decay and solar neutrinos in exposures $>$100\,ton-year.

Selena is particularly efficient in its use of the isotopically-enriched active target since background suppression relies on event-by-event identification of the signal and not self shielding.  Furthermore, our background estimates assume background event rates comparable to those already achieved by kg-scale solid-state detectors for rare event searches, \emph{e.g.}, the M{\footnotesize AJORANA} D{\footnotesize EMON\-STRAT\-OR}~\cite{Majorana:2019nbd}.  Thus, the requirements in the radiopurity of the construction materials and processes, although challenging, are already possible.

Finally, we note that Selena will operate at room temperature, the CMOS APS for its imaging modules will be fabricated with standard commercial CMOS foundry processes, and the aSe deposited following industrial-scale processes developed for the fabrication of medical devices, making the project cost-effective at scale.

We are in the final R\&D stage to validate the aSe/CMOS imaging technology for Selena.  In the next years, we plan to fabricate and test the first detector modules that will serve as the building block for a larger experiment.  The first detector will be a 100-kg Selena ``demonstrator,'' which will directly demonstrate the background suppression capabilities of the experimental technology in an underground environment.  This compact detector (of overall volume $\sim$0.1\,m$^3$) will place the strongest lower limit on the half-life of the neutrinoless $\beta\beta$ decay of $^{82}$Se and detect $\sim$five \emph{pp} solar neutrinos with zero background in a one-year run.  The Selena program will then proceed with an even larger detector.

The white paper is structured as follows. In Sec.~\ref{sec:scientific_motivation}, we present the scientific case for a 10\,ton Selena detector with a 100\,ton-year exposure. In Sec.~\ref{sec:technology}, we describe the detector technology behind Selena, including recent R\&D milestones and the expected performance of a large detector. Finally, in Sec.~\ref{sec:selena_proposed}, we outline the research program for the upcoming years, with a clear path that addresses the biggest challenges for a large detector.

\clearpage

\section{Scientific Motivation}
\label{sec:scientific_motivation}

In the next sections, we present the scientific potential of a 10\,ton Selena detector in three specific areas of neutrino physics and astrophysics. 

\subsection{Neutrinoless \bb\ decay of \ce{$^{82}${Se}}}
\label{sec:selena_bb}

The search for neutrinoless \bb\ decay probes the Dirac-Majorana nature of the neutrino.  This process is forbidden in the lepton-number-conserving Standard Model (SM) and is only allowed if neutrinos are Majorana in nature.  Observation of this process would demonstrate that lepton number is not conserved, with deep implications for grand unification and the matter-antimatter asymmetry of the universe.

When \bb\ decay occurs in a calorimetric detector, the summed energy of the two emitted $\beta$s is observed.  In the SM process involving the emission of two neutrinos, the summed $\beta$ energy follows a continuous spectrum extending up to the $Q$-value of the decay ($Q_{\beta\beta}$).  If no neutrinos are emitted, the two $\beta$s carry away the full decay energy, and a monoenergetic peak would be observed at $Q_{\beta\beta}$.  The shape of the peak is determined by the detector energy resolution.  Thus, the strategy for any neutrinoless \bb\ search involves maximizing exposure to make the amplitude of the peak as large as possible, and minimizing the background rate within a full-width half-maximum of $Q_{\beta\beta}$.

\begin{figure}[t]
  \begin{center}
    \includegraphics[width=\textwidth]{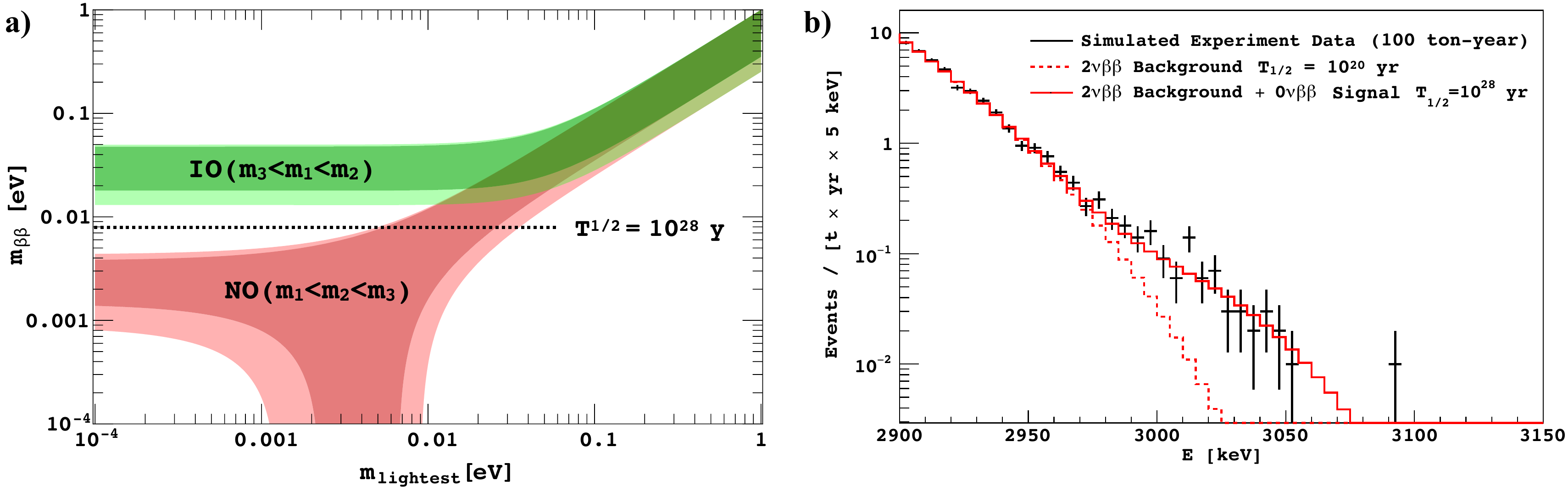}
    \caption{ {\bf a)}~Allowed values for \mbb\ as a function of the mass of the lightest neutrino for different orderings of the neutrino masses~\cite{DellOro:2016gf}. The width of the bands represent the uncertainties in the parameters of the PMNS matrix. The dashed line shows the value of  \mbb\ that corresponds to $\tau_{1/2}=10^{28}$\,y in \ce{^{82}Se}.  {\bf b)}~Predicted \bb -decay spectrum of $^{82}$Se about $Q_{\beta\beta}$ in a 100\,ton-year exposure if $\tau_{1/2}=10^{28}$\,y.}
    \label{fig:science_bb}
  \end{center}
\end{figure}

In a minimal SM extension in which neutrinoless \bb\ decay is mediated by SM neutrinos.  The half-life is inversely proportional to the square of the effective neutrino mass $m_{\beta \beta} = | \sum_i U_{ei}^2 m_i |$, where $U_{ei}$ are PMNS neutrino-mixing matrix elements and $m_i$ is the mass of the $i^{th}$ neutrino mass eigenstate.  Results from neutrino oscillation experiments predict that for the inverted mass ordering (IO: $m_3 < m_1$) there is a lower limit on $m_{\beta \beta}$ at $(m_{\beta\beta}^{min})_{\textrm{IO}} \sim 20$~meV (Fig.~\ref{fig:science_bb}a).  Multiple proposed efforts aim to achieve this sensitivity by operating ton-scale experiments for 5--10 years in several isotopes (namely \ce{^{76}Ge}, \ce{^{100}Mo} and \ce{^{136}Xe})~\cite{Agostini:2021kba, HallmanDNP21}.  For the case of the normal mass ordering (NO: $m_1 < m_3$), most of the parameter space would remain in the range 1\,meV~$<m_{\beta \beta}<(m_{\beta\beta}^{min})_{\textrm{IO}}$

Our initial study on the search for neutrinoless \bb\ decay with a conceptual Selena detector predicts background rates at $Q_{\beta\beta}=3$\,MeV below $6\times10^{-5}$ per keV per ton-year \cite{Chavarria:2016hxk}.  This extremely low background level is possible because of a combination of factors.  First, the high $Q_{\beta\beta}$ of \ce{$^{82}${Se}} is at an energy greater than most backgrounds from primordial $^{238}$U and $^{232}$Th radiocontaminants, which leads to a relatively low raw event rate.  Second, spatio-temporal correlations effectively reject any radioactive decays in the bulk or the surfaces of the imaging modules.  Finally, external $\gamma$-ray backgrounds, which mostly produce single-electron events from Compton scattering or photoelectric absorption, are suppressed by the requirement that the \bb\ signal events have two clearly identified Bragg peaks (this selection retains 50\% of signal events while rejecting 99.9\% of the single-electron background).

The superb background suppression does not require much more than a realistic pixel pitch $<15$\,$\mu$m and an imager uptime close to 100\%, specifications that are within reach of current technology (Sec.~\ref{sec:technology}).  The spectroscopic identification of the neutrinoless \bb\ signal over the two-neutrino \bb\ decay background is a bigger challenge.  Figure~\ref{fig:science_bb}b shows the \bb\ decay spectrum assuming our experimentally-informed best estimate for the energy resolution (red marker in Fig.~\ref{fig:eres}b), and $\tau_{1/2}=10^{28}$\,y for the neutrinoless channel.  Although there is interference from the two-neutrino channel, a limit of $\tau_{1/2}>5\times10^{28}$\,y (90\% C.L.) would be possible in a 100\,ton-year exposure.

In the fortunate event that one of the upcoming ton-scale experiment discovers neutrinoless \bb\ decay, Selena will play a followup role by being the first experiment to confirm the process in the isotope \ce{^{82}Se}, and the first to confirm that it indeed occurs by the emissions of two electrons.  Other possible details of the decay mechanism could be uniquely probed by Selena from the angular correlations between the outgoing electrons~\cite{Ali:2007ec, Deppisch:2020mxv}.

\subsection{Solar neutrino spectroscopy}
\label{sec:selena_solar}

Solar neutrinos have proved to be crucial probes to understand the fundamental properties of neutrinos~\cite{SNO:2011hxd} and stellar astrophysics~\cite{BOREXINO:2018ohr}.  Solar neutrinos are produced by fusion reactions in the sun, which transmute four hydrogen nuclei to one \ce{^{4}He} nucleus.
Neutrinos are emitted both by primary and secondary nuclear processes, with the neutrinos emitted by each reaction labeled a ``species.''  The different neutrino species can be distinguished from their characteristic energies, \emph{i.e.}, by their spectrum.  They provide critical information about the fusion processes in the sun and the fundamental properties of neutrinos as they travel in the dense solar matter and transition into empty space~\cite{Gann:2021ndb}. 

\begin{figure}[t]
  \begin{center}
    \includegraphics[width=0.8\textwidth]{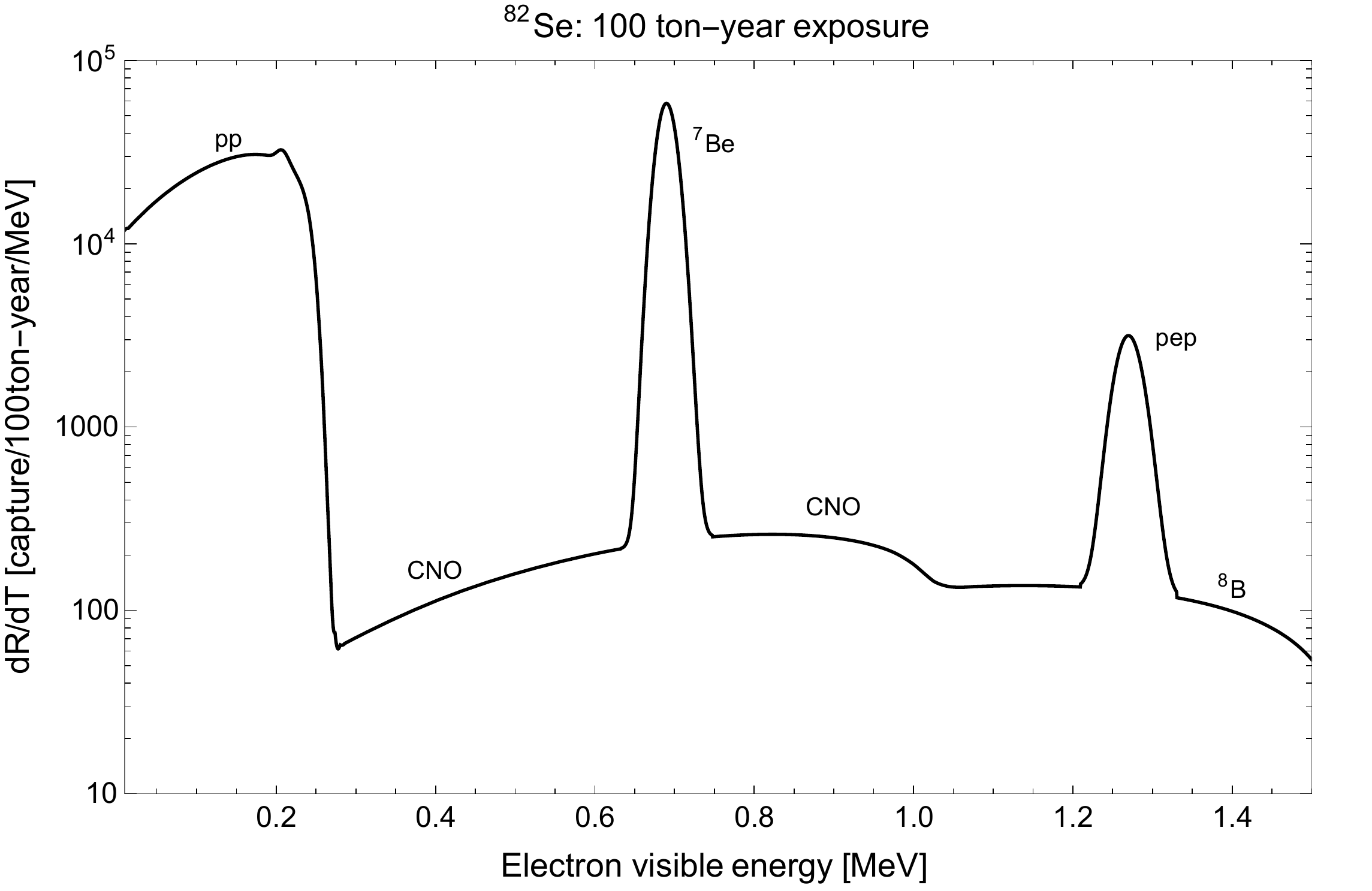}
 \caption{Expected spectrum of the prompt event from solar $\nu_e$ captures. Neutrino species from different solar-fusion reactions are labeled. Capture cross section from Ref.~\cite{PG2022}. Solar neutrino fluxes from Ref.~\cite{Vinyoles:2016djt}. Neutrino oscillation parameters from Ref. \cite{Capozzi:2017ipn}.}
    \label{fig:selena_spectra}
  \end{center}
\end{figure}

Solar $\nu_e$ captures in Selena can be tagged on an event-per-event basis with high efficiency to perform solar $\nu$ spectroscopy concurrently with the search for neutrinoless $\beta\beta$ decay.  The threshold for $\nu_e$ capture of 172\,keV provides sensitivity to all solar $\nu$ species, and leads to the decay sequence shown in Figure~\ref{fig:selena_principle}a. The first step with $\tau_{1/2}=7.2$\,ns is too fast to be distinguished from the electron with energy $E_\nu-172$\,keV emitted following $\nu_e$ capture.  Together they constitute the ``prompt'' event, whose spectrum is shown in Figure~\ref{fig:selena_spectra} with the energy resolution given by the red line in Fig.~\ref{fig:eres}b. The prompt event is followed by a sequence of two decays, which are expected to occur at exactly the same spatial location with time delays of $\sim$5\,min and $\sim$1\,day, respectively.  Although these time separations may seem large, the probability of any two accidental events occurring at exactly the same pixel (\emph{i.e.}, in $\sim$10\,$\mu$g of aSe) is negligible.  Furthermore, any charged particle reaction to produce $^{82m}$Br ($2^-$ state) from $^{82}$Se would have remarkably different prompt event topology.  Additionally, no background isotope has been identified that could mimic the $\nu_e$ capture sequence in time separation, event energies and topologies.  So far, it appears that zero-background solar $\nu$ spectroscopy with a large Selena detector may be possible.

\begin{figure}
\centering

  \subfloat[]{\label{(a)}{\includegraphics[width=.7\linewidth]{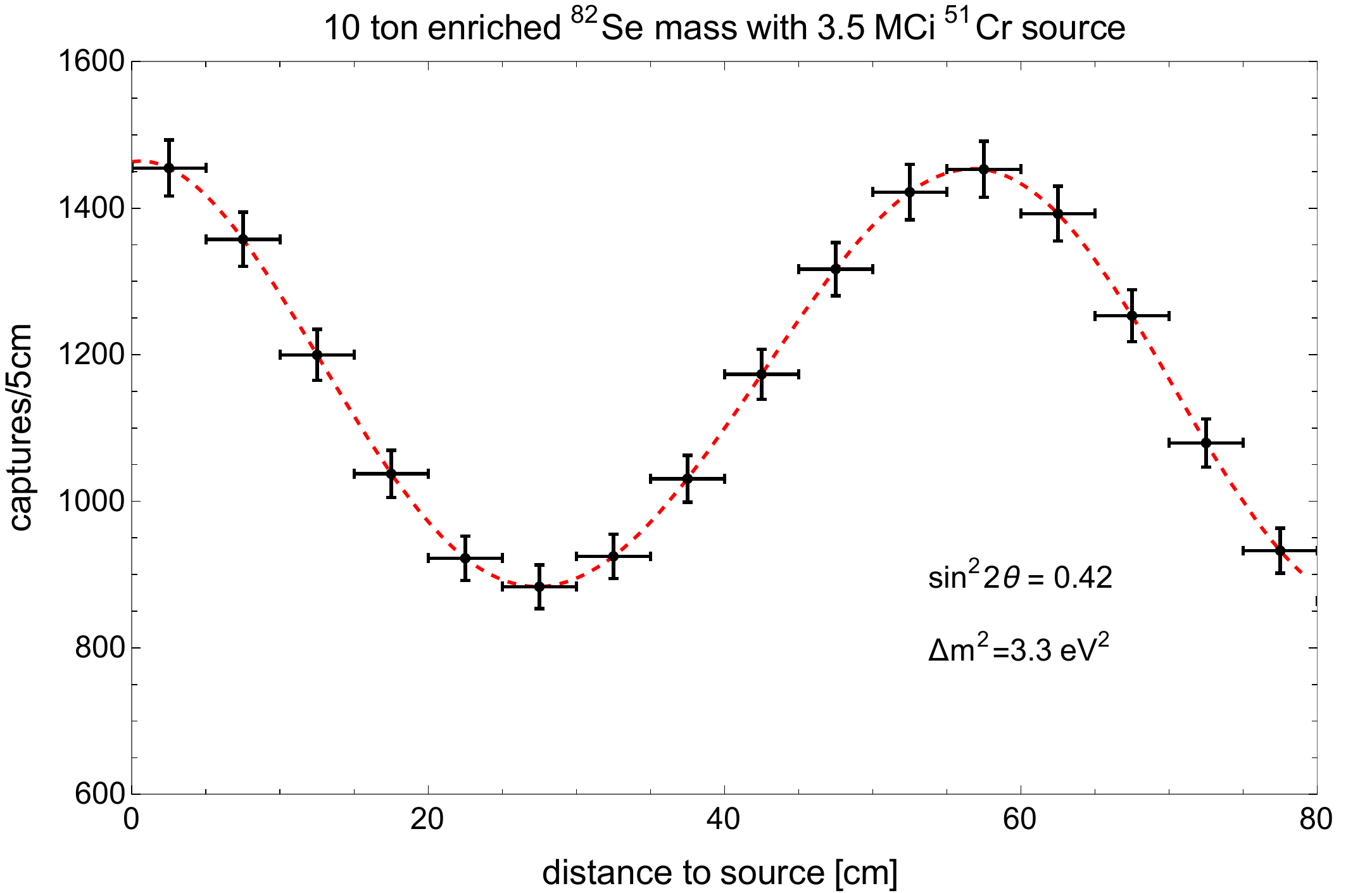}}}
  
  \vspace{0.2cm}
  
  \subfloat[]{\label{(b)}{\includegraphics[width=.7\linewidth]{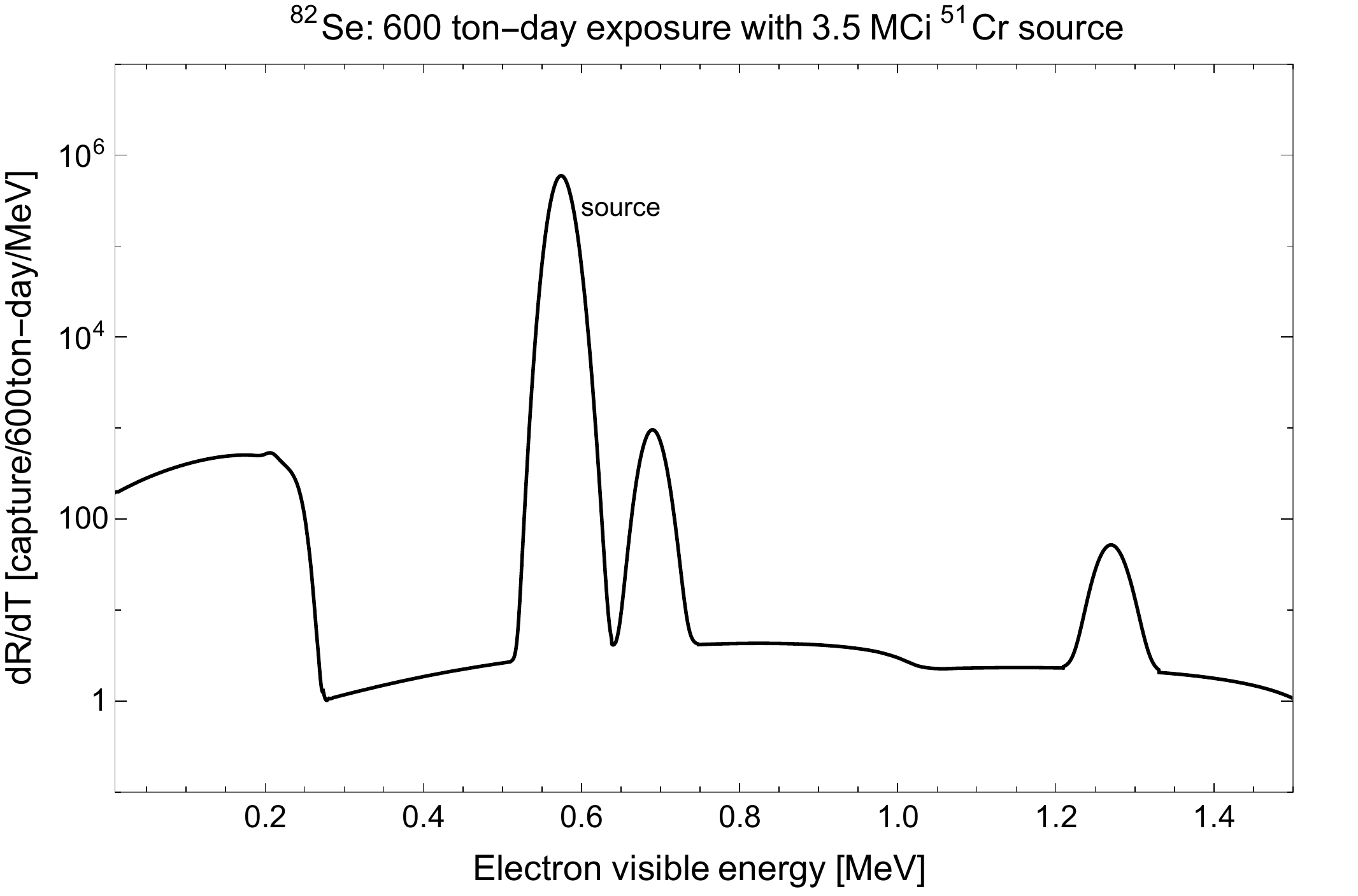}}}

\caption{{\bf a)}~Number of $\nu_e$ captures as a function of distance from an intense \ce{^{51}Cr} source at the center of a 10\,ton Selena detector. Best-fit parameters from Ref.~\cite{Barinov:2021asz}. {\bf b)}~Expected spectrum of prompt events from \ce{^{51}Cr} $\nu_e$ captures over the solar neutrino background.}
\label{fig:selena_source}
\end{figure}

\begin{table}[h]
  \begin{center}
    \begin{tabular}{lccc}
      Species &  $E$ range [keV] & $N$ in 100 ton-year & $1/\sqrt{N}$ \\
      \hline
	  {\it pp} &	29--278	& 5359 & 1.4\% \\
	  $^7$Be & 665--775  & 1814 & 2.3\% \\
	  {\it pep} & 1230--1360 & 130 & 8.8\% \\
	  CNO & 278--655, 785--1220& 136 & 8.6\% \\
	  $^8$B & (1.5-15)$\times10^3$ & 209 & 6.9\% \\
    \end{tabular}
    \caption{Expected results for the capture rate of solar neutrinos from a counting experiment. $N$ is the expected number of $\nu_e$ captures for the different species in the specified energy ($E$) range. The last column provides an estimate of the statistical uncertainty in the measured rate.}
    \label{tab:selena_solar}
  \end{center}
\end{table}

Table~\ref{tab:selena_solar} presents the capture rates for different solar neutrino species obtained by integrating Fig.~\ref{fig:selena_spectra} in the specified energy regions. Neutrino charged current cross sections on $^{82}$Se have been taken from Ref.~\cite{PG2022}. With a 100\,ton-year exposure, Selena is expected to measure the $pp$ rate to $\sim$1\% for a strong constraint on the neutrino luminosity\footnote{The uncertainty in the $pp$ flux will be dominated by the uncertainty in the $\nu_e$ capture cross section. For example, calibration to the $\nu$-$e$ elastic scattering cross section from a comparison between the $^7$Be neutrino rates measured by Selena and Borexino~\cite{BOREXINO:2018ohr} would contribute a 3.5\% uncertainty.}, the $pep$ rate to $\sim$8\% to probe the onset of the Mikheyev-Smirnov-Wolfenstein (MSW) effect for solar neutrinos~\cite{Mikheev:1987qk}, and the CNO-cycle rate to $\sim$10\% to demonstrate the presence of the CNO-cycle in the Sun with 5$\sigma$ significance and to resolve the solar metallicity problem~\cite{Serenelli:2009yc}.

\begin{figure}[t]
  \begin{center}
    \includegraphics[width=0.6\textwidth]{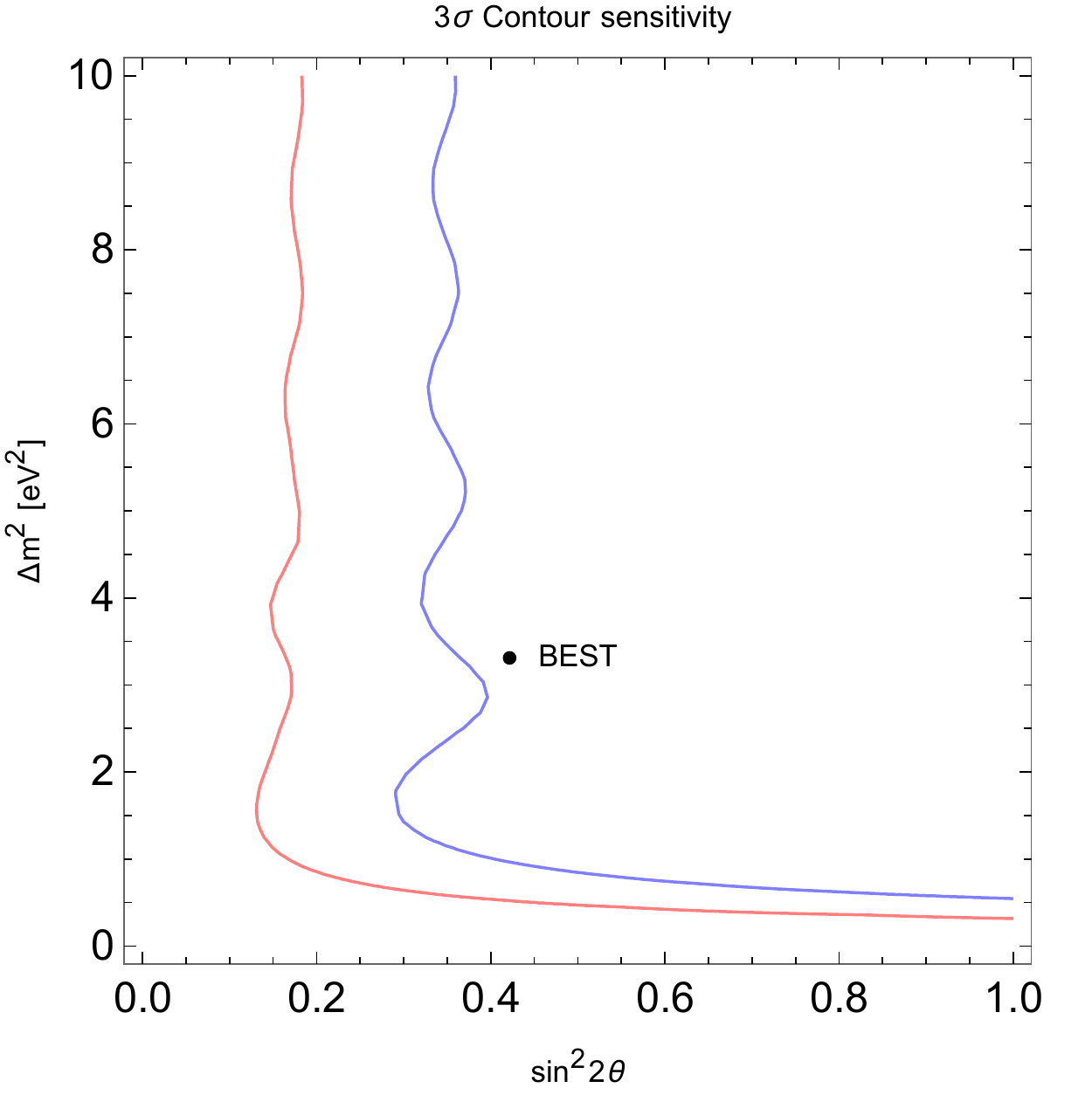}
 \caption{Sensitivity to oscillation parameters from an intense \ce{^{51}Cr} source at the center of a 10\,ton Selena detector. The best-fit from Ref. \cite{Barinov:2021asz} is shown for comparison. Blue line: only the total rate of the neutrino capture line within the detector volume is considered. Red line: the oscillation pattern is exploited to enhance sensitivity to neutrino parameters.}
    \label{fig:source_sensitivity}
  \end{center}
\end{figure}

The shape of the energy spectrum of neutrino lines is modified by thermal effects. On the low energy side of the line, the dominant effect is first-order Doppler broadening caused by nuclear motion. On the high energy side of the line, the dominant effect is exponential broadening caused by center of energy electron-nucleus. The asymmetric broadening causes a shift in the mean energy of the solar neutrino lines. The $^7$Be electron-capture decay line is expected to be shifted by 1.3\,keV, relative to the neutrino line in the laboratory~\cite{Bahcall:1994cf}, directly correlated with the solar core temperature.  There is a corresponding shift of the \emph{pep} line by 7.6\,keV, much larger due to the lighter nuclei  involved in the electron capture~\cite{PG2022}.  We expect a Selena experiment with a 100\,ton-year exposure to measure, with a dedicated energy calibration better than 1\%, these line shifts with high statistical significance for a $\sim$30\% estimate of the solar core temperature, in what would be the first ever direct measurement of the Solar core temperature. 

\subsection{Sterile neutrinos}
\label{sec:selena_oscillations}

Recent results from the BEST experiment~\cite{Barinov:2021asz} confirm the longstanding ``gallium anomaly''~\cite{Kostensalo:2019vmv}, where radiochemical detectors based on the target isotope \ce{^{71}Ga} observe a $\nu_e$ capture rate that is $80\pm5$\% of the expected from electron-capture sources.  These radioactive sources (\ce{^{51}Cr}, \ce{^{37}Ar}) emit $\sim$0.8\,MeV neutrinos and were deployed at short distances (typically $<$1\,m) from the detector.  This discrepancy has been interpreted as $\nu_e$ oscillations into a sterile neutrinos ($\nu_s$)~\cite{Laveder:2007zz}.  Since the BEST detector saw no difference between the capture rate in its separate inner and outer volumes, if the oscillation hypothesis is correct, the oscillation length must be $<$2\,m.

If a 3.5\,MCi \ce{^{51}Cr} source were to be placed in the center of a 10\,ton Selena detector, in a similar configuration to the BEST experiment, $\sim 2.3\times 10^4$\, $\nu_e$ capture events would be tagged with high efficiency\footnote{For reference, consider the proposed LENS-Sterile experiment~\cite{Grieb:2006mp}, which features a segmented real-time calorimetric $\nu_e$ detector with a response that is qualitatively similar to Selena's.}. This estimation takes into account a source transportation time after irradiation of 4 days and an exposure of 60 days. Calculations method is similar to the one discussed in Ref.~\cite{Ianni1999}.
Furthermore, Selena would be able to resolve the spatial location of every $\nu_e$ capture to search for oscillations with lengths as small as $\sim$0.1\,m, limited only by the physical size of the radioactive source.  In terms of neutrino oscillation parameters, this corresponds to probing the short-baseline $\nu_s$ oscillation hypothesis in the range 1\,eV$^2<\Delta m^2<$10\,eV$^2$.  Fig.~\ref{fig:selena_source} shows the simulated result where the gallium anomaly is explained by $\nu_e \rightarrow \nu_s$ oscillations. Selena would be able to clearly resolve the oscillation pattern to conclusively test the hypothesis. The same figure shows the expected spectrum for a 10\,ton Selena detector with the source at the center. In order to probe the sensitivity of this detector to neutrino oscillation parameters, we compared the predicted \nue\ capture rate with and without the source. The result is shown in Fig.~\ref{fig:source_sensitivity}, where the best-fit from BEST from Ref.~\cite{Barinov:2021asz} is shown for comparison. Inclusion of the oscillation pattern enhances significantly the potential of Selena in this framework.

\section{Detector technology}
\label{sec:technology}

Amorphous selenium is a well-established photoconductor commonly used as an X-ray converter in flat panel imagers employed in the medical industry for radiographic and fluoroscopic digital imaging~\cite{Zhao:1995fx,rowlands}.  Incident radiation creates electron-hole pairs in the target material and these carriers are separated by a strong electric field (typically tens of V/$\mu$m) applied across the detector.  The charge carriers drift towards the electrode where they induce a measurable signal proportional to the amount of charge generated.  Current medical devices use thin-film transistor (TFT) technology to measure the charge in each pixel, resulting in $\sim$1000\,$e^-$ of noise per pixel, making this style detector insufficient for single electron detection and spectroscopy\footnote{For example, a 50\,keV X ray generates a signal of 1000\,$e^-$ at an electric field of 10\,V/$\mu$m.}.  Thus, in Ref.~\cite{Chavarria:2016hxk}, we proposed to read out the ionization charge from aSe with a lower noise CMOS APS to resolve individual electron tracks. Details of the proposed detector for the Selena experiment are discussed in the following sections.

\subsection{Imaging Capabilities}
\label{sec:detector_imager}

\begin{figure}[t!]
 	\begin{center}
 		\includegraphics[width=\textwidth]{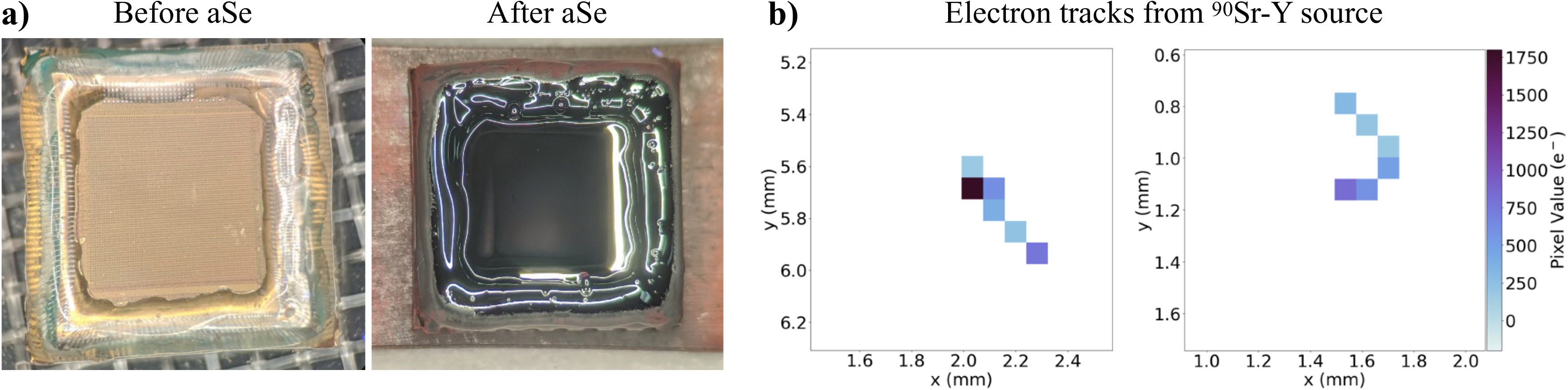}
 		\caption{{\bf a)} \TMIIm CMOS APS before and after aSe deposition. {\bf b)} Electron tracks observed with our first hybrid aSe/CMOS device operated at a drift electric field of 4\,V/$\mu$m at room temperature.}
 		\label{fig:selena_topmetal}
 	\end{center}
 \end{figure}

We recently demonstrated the imaging of single electron tracks with a hybrid aSe/CMOS device for the first time.
The device was based on the \TMIIm CMOS APS~\cite{An:2015oba}, which features a rectangular array of 72$\times$72 pixels with 83\,$\mu$m pitch.
We achieved a pixel noise of 23\,$e^-$ with the sensor coupled to a 500\,$\mu$m thick aSe target layer.
Fig.~\ref{fig:selena_topmetal}a shows photographs of the device before and after aSe deposition. Fig.~\ref{fig:selena_topmetal}b shows electron tracks from a \ce{^{90}Sr} $\beta$ source obtained with the device operated at high voltage.

The Selena experiment will consist of much larger imagers made from $\sim 5$\,mm-thick amorphous \ce{^{82}Se} layers coupled to an array of $\sim 5$\, cm$^2$ CMOS active pixel arrays tiled on a 300-mm diameter silicon wafers.
The aSe layers are sufficiently thick to fully-contain the mm-long \bb\ decay tracks with total energy of $Q_{\beta\beta}$.
Each module is made from two back-to-back imagers that share a common electrode that provides the high voltage (HV). A detector is built from towers of these modules inside a radio-pure lead shield in an underground laboratory to minimize background.

\begin{figure}[t!]
  \centering
  \includegraphics[width=\textwidth]{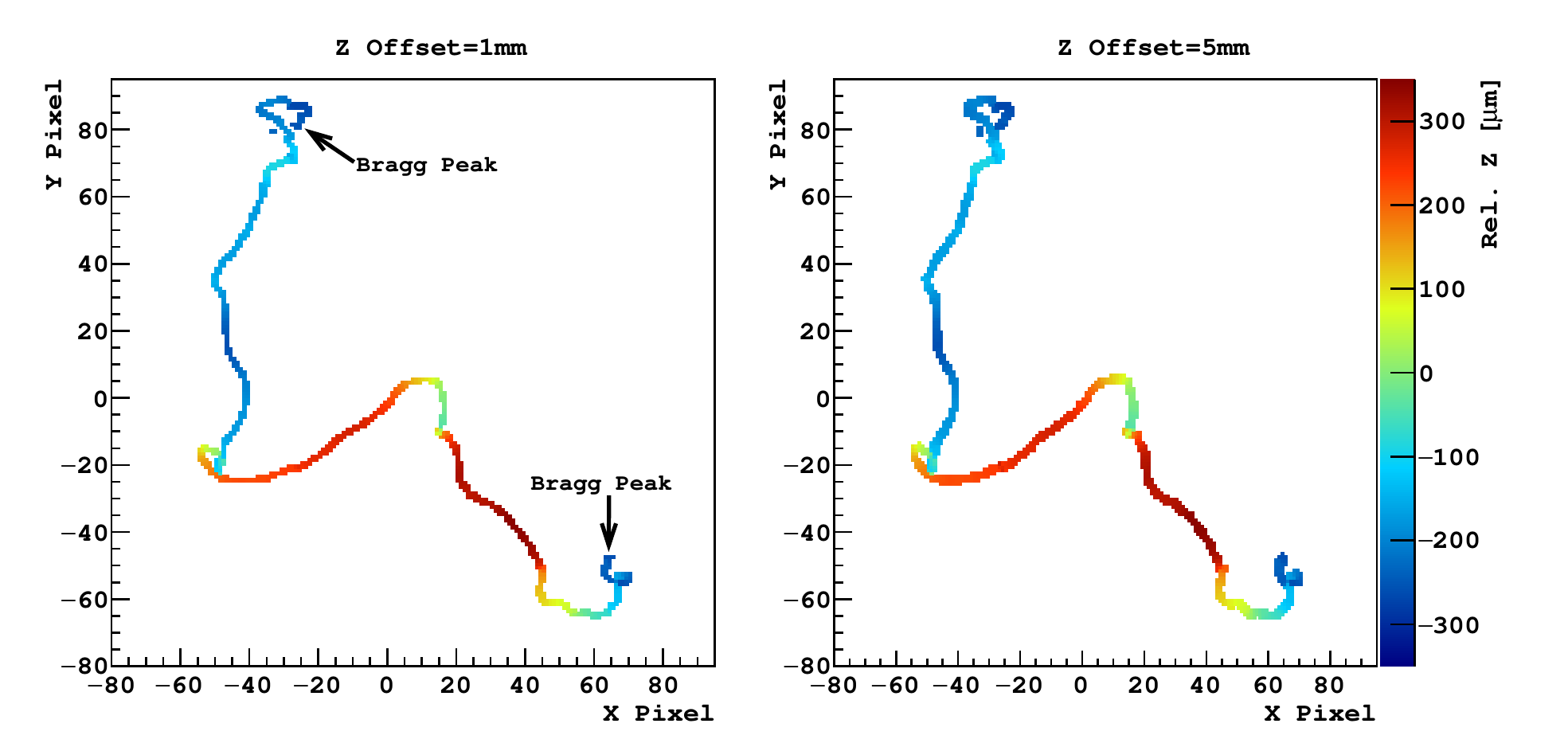}
  \caption{Simulated \bb\ decay event with total energy of $Q_{\beta\beta}$ in an imager with {\rm 10$\times$10 $\mu$m$^2$} pixel size. The color axis corresponds to the relative $z$ position. In the left plot, we simulated lateral diffusion of charge with an overall offset on the $z$ position of 1\,mm, while the right plot has an overall $z$ offset of 5mm. The Bragg peaks are identified and labeled.}
  \label{fig:TracksNoField}
\end{figure}

The CMOS APS will feature standard parameters for an imaging device: pixels $10 \times 10$ \um$^2$ in size, a pixel readout noise of 10\,$e^-$ and a pixel dynamic range of 70\,dB. 
Each pixel will be instrumented to measure the total integrated charge collected by a pixel and the time of arrival of the charge.
The integrated charge provides an estimate of the deposited energy, while the time of arrival will be used to reconstruct the relative $z$ coordinate of the energy depositions for high-resolution 3D imaging.
Considering the drift speed of the charge carriers, a 5\,ns time resolution would correspond to a resolution in $z$ of 10\,$\mu$m.
We will also infer the absolute $z$ position of the events from the lateral spread of charge on the pixel array. 
Fig.~\ref{fig:TracksNoField} shows the differences caused by lateral charge diffusion of the same simulated $\beta\beta$ decay event at different ($1$\,mm vs. $5$\,mm) depths in the detector.

\subsection{Energy Response}
\label{sec:detector_energy_response}

\begin{figure}[t!]
  \centering
  \includegraphics[width=\textwidth]{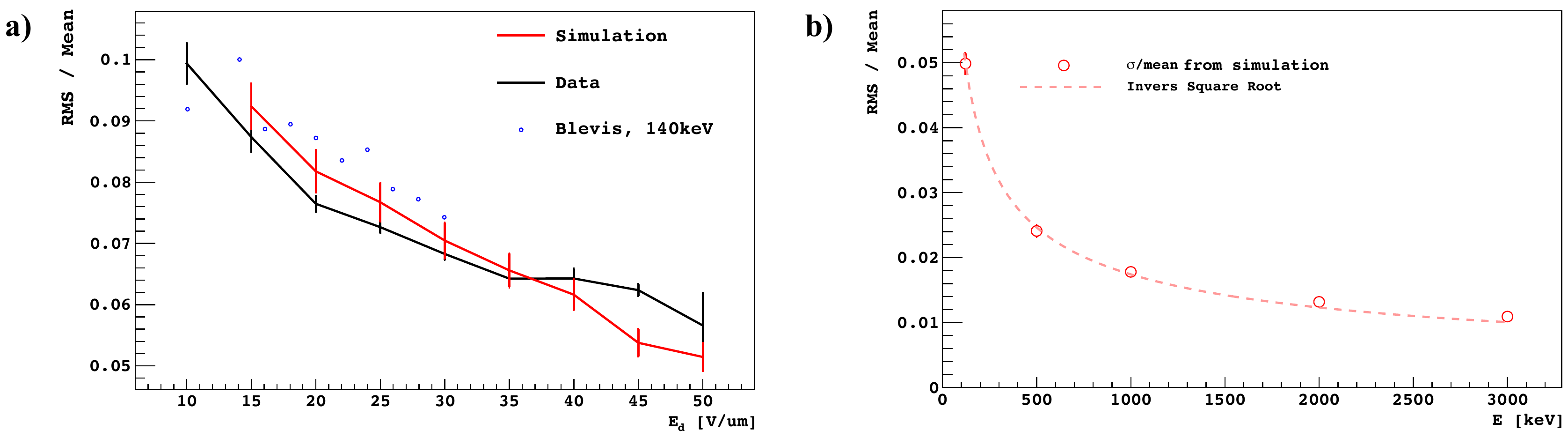}
  \caption{{\bf a)} Fractional energy resolution of the 122\,keV \gr\ line from \ce{^{57}Co} as a function of drift electric field in aSe. Our full detector simulation that includes our recombination model (red line) is in good agreement with the data. {\bf b)} Extrapolation with our detector simulation of the fractional energy resolution up to $Q_{\beta\beta}$ at a drift electric field of 50\,V/\um .}
  \label{fig:eres}
\end{figure}

The energy resolution at $Q_{\beta\beta}$ determines the capability of Selena to spectrally resolve the neutrinoless $\beta\beta$ decay signal over the two-neutrino $\beta\beta$ decay background spectrum.
In Ref.~\cite{Li:2020ryk}, we reported our measurement of the response of aSe to 122\,keV $\gamma$ rays from a $^{57}$Co radioactive source as a function of the electric field in the aSe.
Fig~\ref{fig:eres}a shows the measured energy resolution compared to our full detector simulation, which includes particle tracking, all details of the charge generation and transport, and an electronics simulation.
The data is best explained by a recombination probability that has a strong dependence on the ionization density along the electron tracks, \emph{i.e.}, by a charge yield along that has an inverse relation with $dE/dx$.
This leads to an energy resolution that is dominated by fluctuations in $dE/dx$ and not charge-carrier statistics.
Fortunately, the imaging capabilities of Selena will allow for a local measurement of $dE/dx$ along the electron tracks from which we can estimate the energy by correcting for the varying charge yield.
From our latest simulations based on our recombination model, we expect an energy resolution at $Q_{\beta\beta}$ of 1.1\% RMS after correcting for the varying charge yield.
Although further improvements may be possible, we use the extrapolation shown in Fig.~\ref{fig:eres}b to build the science case in Sec.~\ref{sec:scientific_motivation}.

The calibration of the energy response of Selena is straightforward since the detector identifies individual electron tracks from single-scattering events.
Photoelectric absorption of monoenergetic $\gamma$ rays leads to a monoenergetic sample of electron tracks that are easily distinguishable from the continuous spectrum of electrons from Compton scattering since their energies are significantly above the Compton edge.
By exposing the Selena modules with $\gamma$ ray sources of various energies up to $Q_{\beta\beta}$, we would be able to calibrate with high accuracy the entire energy region of interest for solar neutrino spectroscopy and \bb\ decay.
Furthermore, the data will contain not only the total ionization signal of the events but high resolution 3D images of the ionization tracks.
With sufficient statistics, we should be able to experimentally determine the charge yield as a function of $dE/dx$ to validate our recombination model and optimize the energy resolution.

\subsection{Background Sources and Rejection}
\label{sec:detector_backgrounds}

In Ref.~\cite{Chavarria:2016hxk}, we studied by simulation how the imaging capabilities of a large Selena detector could be employed to suppress backgrounds from natural radioactivity in the search for neutrinoless \bb\ decay.
We performed an extensive survey of radiogenic and cosmogenic background sources considering radiocontaminants in the bulk and on the surfaces of the modules, and an external \gr\ flux.
We note that since only a very small fraction of $\gamma$ rays from \ura\ and \tho\ decay chains can produce single electrons with energies of $Q_{\beta\beta}$, a significant contribution to the relevant \gr\ flux comes from cosmogenic isotopes activated \emph{in-situ} and deexcitation of nuclei following neutron capture.
This leads to unusual constraints in the construction materials since it introduces considerations (\emph{e.g.}, activation and neutron capture cross sections) beyond material radiopurity.
We argued that classification of events by spatio-temporal correlations could suppress backgrounds originating in the bulk or on the surfaces of the modules to a negligible level.
A $10^3$ reduction in background from the external \gr\ flux was achieved by discarding events without two Bragg peaks identified by an automated algorithm.
This could perhaps be improved by leveraging on more advanced machine learning techniques~\cite{NEXT:2020jmz}.
The final value of $6\times10^{-5}$ per keV per ton-year would allow us to perform background-free \bb\ decay spectroscopy in 100\,ton-year.

The number of accidentals that mimic the ``triple sequence'' used to identify \nue\ captures (Fig.~\ref{fig:selena_principle}a) is $<10^{-4}$ in 100 ton-year in an underground detector, whose background rate is dominated by two-neutrino \bb\ decay.
Charged particle reactions to produce \ce{^{82m}Br} from \ce{^{82}Se}\textemdash the traditional background for radiochemical experiments\textemdash have a prompt event topology that is easily distinguishable from a fast electron.
A more likely background for the production of \ce{^{82m}Br} would be from neutron capture on trace \ce{^{81}Br} in the aSe, where the deexcitation $\gamma$ ray only interacts once very close to the decay site to mimic the prompt event.
However, the contamination of Br in aSe is not known at this time.
The dominant background will likely come from neutron captures on \ce{^{82,80,78}Se}, which lead to sequences of three or four decays that emit fast electrons and $\gamma$ rays.
However, they can all be suppressed by appropriate selections on event topology, time separation and decay energies.
Finally, we performed a survey of isotopes that can be cosmogenically activated in \ce{^{82}Se} and could not identify any that could be produced at any significant rates and could mimic the triple sequence closely enough in topology and time separation.
Thus, so far, it appears that zero-background solar neutrino spectroscopy is possible in a 100 ton-year exposure with Selena.

\section{The Selena Research program}
\label{sec:selena_proposed}

In the next sections, we describe technical details and ongoing developments to realize the Selena Neutrino Experiment.

\subsection{Pixel structure}
\label{sec:selena_sensorRD}

The pixel structure of Selena will be optimized for \emph{i)}~charge collection efficiency $>$99.9\%, \emph{ii)}~noise on charge sensing per pixel of $\sim$10\,$e^-$, \emph{iii)}~time-of-arrival estimate of the charge to the pixel with an accuracy of 5\,ns, \emph{iv)}~pixel pitch of 10\,$\mu$m, and \emph{v)}~minimal power consumption for the readout of large pixel arrays.

Each pixel in the CMOS charge sensor will focus the charge coming from the aSe layer to the collection electrode and capture the charge signal with circuitry embedded in the pixel.  The outputs of the pixel circuitry, which correspond to the charge amplitude and time of arrival, will be driven out of the pixel array and recorded by external readout.  The \TMIIm~\cite{An:2015oba} realized most of these goals but was not specifically optimized for aSe operations, which resulted in sub-optimal performance.

\begin{figure}[t!]
  \begin{center}
    \includegraphics[width=\textwidth]{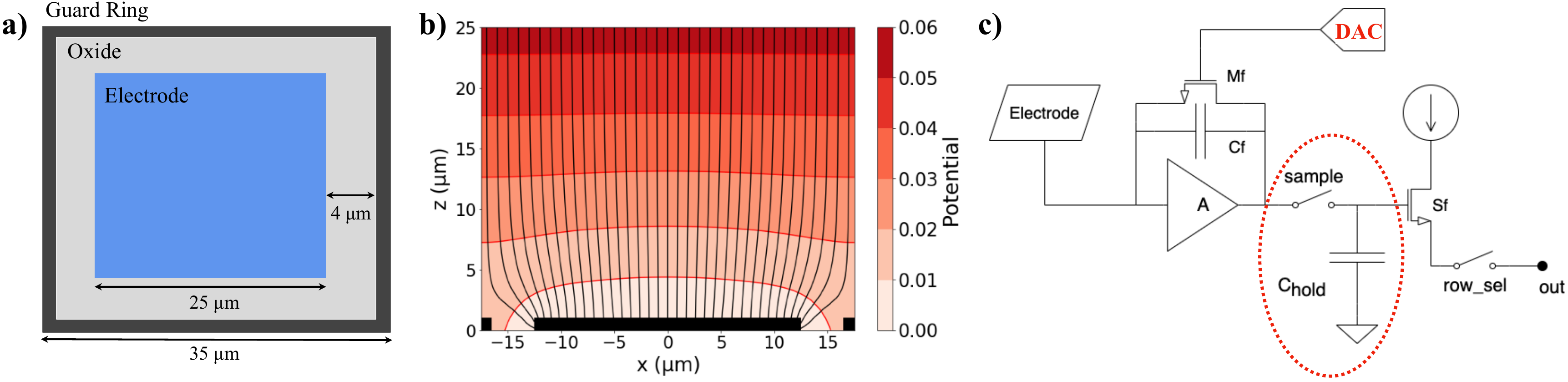}
    \caption{{\bf a)} Proposed pixel structure for the next CMOS APS iteration. {\bf b)} Electrostatic simulation results (electric field lines and equipotential lines) for the proposed pixel structure. Collection efficiency of $\sim$100\% is achieved with a guard ring potential of 1.5\% of the HV. {\bf c)} Schematic of the pixel circuitry of the proposed design. Parts in red (a sample/hold for the pixel value, and a DAC to set the feedback resistance of the pixels individually) are new additions to the baseline \TMIIm design.}
    \label{fig:selena_improve}
  \end{center}
\end{figure}

To achieve maximum charge collection efficiency and minimize inter-pixel cross talk, we plan to implement a charge collection and focusing electrode structure in the topmost metal layer of CMOS as shown in Fig.~\ref{fig:selena_improve}a.  The center electrode collects charge and transfers it to the input of the CSA.  The guard ring, which surrounds the center electrode and forms a cross-pattern between pixels, is biased to focus the charge to the electrode and to isolate pixels from each other.  The results from electrostatic simulations shown in Fig.~\ref{fig:selena_improve}b suggest that 100\% collection efficiency is possible with a guard ring potential of only 1.5\% of the drift HV.

We also plan to improve the in-pixel circuitry by implementing, per pixel, a DAC tunable CSA feedback, a sample-hold structure of the CSA output, and a time-of-arrival circuit (Fig.~\ref{fig:selena_improve}c).  The per-pixel tunable DAC allows fine-tuning the time constant of the CSA in each pixel, to counter the process variation from pixel to pixel and ensure sufficient signal gain and retention.  Further, a sample-hold structure shall be added to the output of the CSA to act upon a trigger to capture the signal and wait for readout sometime later.  The sample switch is opened by a discriminator on the pixel (not shown but present in the current \TMIIm design) when its value crosses a programmable threshold set by a DAC.  These two additions are expected to significantly improve the uniformity and overall resolution of the sensor.  The same discriminator circuit can be used to sample a clock to provide the time-of-arrival of the charge to the pixel\footnote{An existing version of \emph{Topmetal}, \emph{Topmetal-M}, already features time-of-arrival measurement capabilities~\cite{REN2020164557}.}.

\subsection{Large-area detector module}
\label{sec:selena_largesensor}

The Selena detector modules will consist of a large array of stand-alone $2\times2.5$\,cm$^2$ CMOS APSs on a 300\,mm-diameter wafer.  To efficiently contain electron tracks in the energy region of interest in a single module, the aSe layer must be  $\sim 5$\,mm thick, which correspond to a $^{82}$Se target mass of 2\,kg per module.  A significant challenge will be the application of the HV to the modules.  Unlike test devices, where the HV electrode is fully exposed on top, Selena modules will feature face-to-face devices where the HV comes in from the side.  This becomes challenging for a design that maximizes the thickness of the aSe layer.

\begin{figure}
  \centering
  \includegraphics[width=0.9\linewidth]{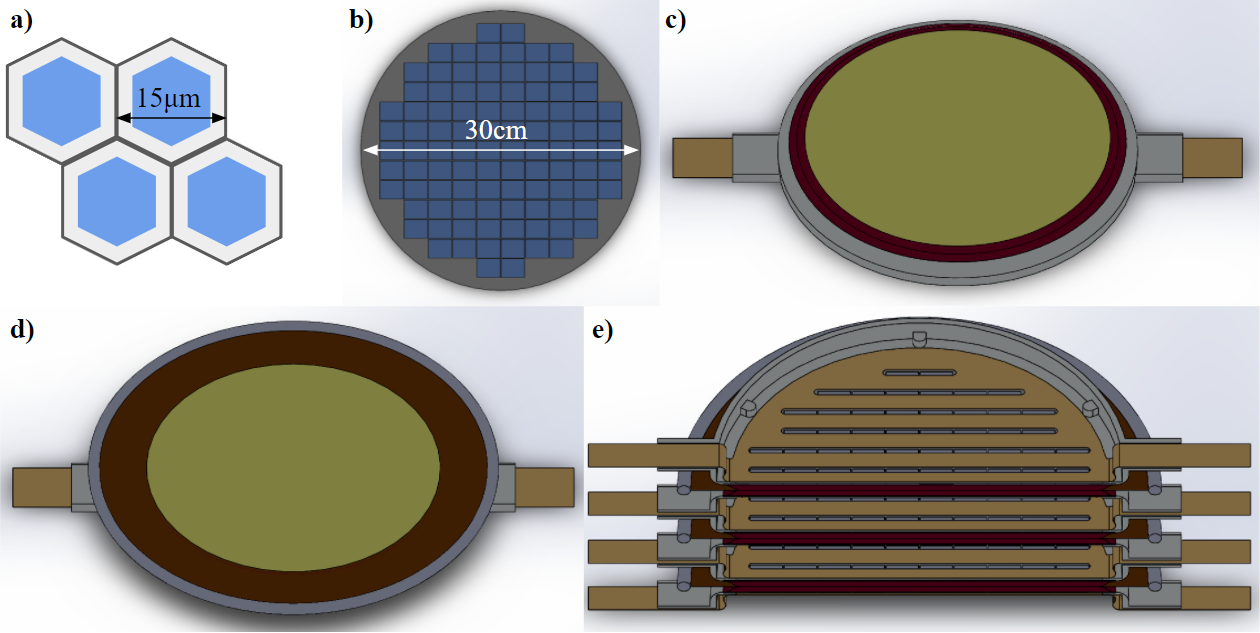}
  \caption{Conceptual design of the large-area detector module. {\bf a)}~Hexagonal pixel array. {\bf b)}~Top surface of the 300-mm diameter wafer. Blue rectangles represent CMOS APSs and gray areas are grounded with post-fabrication coating. {\bf c)}~Single module with gold electrode (gold) deposited on aSe (dark red) on the top wafer surface shown in (b), supporting frame (light gray), and flexible PCB (khaki). {\bf d)}~HV ring placed on top of (c). The brown aperture is conductive Kapton. The gray ring provides mechanical strength and a rounded edge. {\bf e)}~Cross section of a stack of three face-to-face module pairs. Slots on the flexible PCBs are for connecting to TSVs.}
  \label{fig:selena_module}
\end{figure}

We will approach this challenge with the conceptual design shown in Fig.~\ref{fig:selena_module}.  The pixel pattern will be hexagonal (Fig.~\ref{fig:selena_module}a) instead of square to reduce the maximum field near electrode corners while maintaining the same hole collection efficiency.  The profile of the pixel array will remain rectangular.  To maximize the active pixel area and simplify the packaging, we will bring the CMOS APS connections to the backside of the wafer with through-silicon vias (TSVs).  The front side of the wafer (Fig.~\ref{fig:selena_module}b) will be held at ground potential, by the active pixel arrays, and by a conductive coating above the gaps between CMOS APSs and the edges of the wafer deposited post-fabrication.  A thin flexible PCB will be attached to the backside of the wafer and wire-bonded to the TSVs (Fig.~\ref{fig:selena_module}e).  The PCB will deliver power and control signals to CMOS APSs and bring digitized signals outside the Selena module.  The CMOS APS design will feature power regulation and complete digital control, thus no external electric components will be mounted on the PCB for power distribution, chip control, or signal multiplexing.  The PCB will be protected from HV by a conductive frame around the wafer (Fig.~\ref{fig:selena_module}c).  The frame can be metal, or insulator with conductive coating; the later is preferred to reduce radioactive contamination and neutron capture near the modules.  The wafer-PCB-frame assembly will be sent out for aSe deposition.  Making connections prior to aSe deposition avoids mechanically or thermally stressing the aSe film, which is critical for handling high electric field.  A thin layer of gold will be deposited on top of aSe as the HV electrode with a 2\,cm margin to the edge of the wafer.  A conductive Kapton film aperture expended by a conductive ring will deliver HV to the gold electrode (Fig.~\ref{fig:selena_module}d). The aperture in the center will prevent electrons crossing the top of the aSe film from losing energy in the Kapton film. The ring provides mechanical support and a rounded edge far away from ground potential. The film will be gently sandwiched by two Selena modules face-to-face, constituting one basic detector unit.  Finally, multiple pairs of modules will be stacked together as a tower (Fig.~\ref{fig:selena_module}e, the overall support structure is not shown).  The entire assembly will be hosted in vacuum at room temperature. Additional heat dissipation structures attached to the backside of the wafer may be necessary.

\subsection{The Selena 100\,kg demonstrator}
\label{sec:selena_100kg}

A ``demonstrator'' detector made from 50 large-area modules and a target mass of 100\,kg of enriched \ce{$^{82}$Se} will be a stepping stone in the Selena program.
The demonstrator will be designed to minimize backgrounds from natural radioactivity and deployed underground (\emph{e.g.,} at the Laboratorio Subterr\'aneo de Canfranc (LSC) in Spain). Its main goal will be to demonstrate the technology (including scalability), performance and background rejection capabilities necessary for a large-scale Selena detector.

The fabrication of the demonstrator will require the enrichment and deposition of aSe at scale.
Our industrial partner, Hologic, Inc., deposits $\sim$1\,ton of aSe per year for their commercial imagers.
We will setup a dedicated aSe deposition facility for Selena, which will a recovery process to minimize the waste of aSe enriched in \ce{$^{82}$Se}.

Enrichment of \ce{$^{82}$Se} is commonly achieved by centrifugation of gaseous selenium hexafluoride (\ce{SeF_6}), an inefficient process that would dominate the cost of the demonstrator.
A more attractive approach is isotopic enrichment by cryogenic distillation of hydrogen selenide (\ce{H_2Se}), for which a hint of isotopic separation of \ce{^{82}Se} was observed with a meter-long column~\cite{Mills:1988tx}. 
We will explore this possibility as part of the Selena program.
We note the synergies with other efforts in our community, \emph{e.g.,} the Aria project, which aims to remove \ce{$^{39}$Ar} from natural argon by distillation~\cite{DarkSide-20k:2021nia}.

The demonstrator will provide the opportunity to develop a calibration strategy with \gr\ sources for Selena, and will also acquire a large sample of two-neutrino \bb\ decay events.
After one year of operation, the Selena demonstrator will place an lower limit on the neutrinoless decay half-life of \ce{^{82}Se} $\tau_{1/2}>2\times10^{26}$\,y (90\% C.L.), and detect $\sim 5$ \emph{pp} solar neutrinos with zero background.

\clearpage
\bibliography{myrefs}

\end{document}